\documentclass[11pt]{article}
\usepackage{graphicx}
\usepackage{amsmath}
\usepackage{amssymb}

\title{On Conformal Field Theory of SLE$(\kappa, \rho)$}
\author{}
\date{}

\begin{document}

\maketitle

\centerline{Kalle Kyt\"ol\"a}
\centerline{\small \texttt{kalle.kytola@helsinki.fi}}  

\bigskip

\centerline{\large Department of Mathematics and Statistics, P.O. Box 68}
\centerline{FIN-00014 University of Helsinki, Finland.} 

\bigskip

\newcommand{\bC} {\mathbb{C}}
\newcommand{\bR} {\mathbb{R}}
\newcommand{\bZ} {\mathbb{Z}}
\newcommand{\bN} {\mathbb{N}}
\newcommand{\bD} {\mathbb{D}}
\newcommand{\bH} {\mathbb{H}}
\newcommand{\bS} {\mathbb{S}}
\newcommand{\sL} {\mathcal{L}}
\newcommand{\sD} {\mathcal{D}}
\newcommand{\sF} {\mathcal{F}}
\newcommand{\ud} {\mathrm{d}}
\newcommand{\bn} {\mathbf{n}}
\newcommand{\Order} {\mathcal{O}}
\newcommand{\order} {o}
\newcommand{\bdry} {\partial}
\newcommand{\Oper} {\mathcal{O}}
\newcommand{\unit} {\mathbf{1}}
\newcommand{\Bra} {\Big\langle}
\newcommand{\Ket} {\Big\rangle}
\newcommand{\bra} {\langle}
\newcommand{\ket} {\rangle}
\newcommand{\prob} {\mathbb{P}}
\newcommand{\expect} {\mathbb{E}}
\newcommand{\cconj} {\overline}
\newcommand{\im} {\Im \textrm{m }}
\newcommand{\re} {\Re \textrm{e }}
\newcommand{\half} {\frac{1}{2}}

\newtheorem{theorem}{Theorem}
\newtheorem{lemma}{Lemma}
\newtheorem{corollary}{Corollary}
\newcommand{\proof} {\emph{Proof: } }
\newcommand{\QED} {$\square$}

\begin{abstract}
SLE$(\kappa; \underline{\rho})$, a generalization of chordal
Schramm-L\"owner evolution (SLE), is discussed from the point of view
of statistical mechanics and conformal field theory (CFT). Certain
ratios of CFT correlation functions are shown to be martingales.
The interpretation is that SLE$(\kappa; \underline{\rho})$
describes an interface in a statistical mechanics model whose
boundary conditions are created in the Coulomb gas formalism
by vertex operators with charges
$\alpha_j = \frac{\rho_j}{2 \sqrt{\kappa}}$. The total charge
vanishes and therefore the partition function has a simple
product form. We also suggest a generalization of
SLE$(\kappa; \underline{\rho})$.
\end{abstract}


\section{Introduction}
Schramm-L\"owner evolutions (SLEs) are conformally invariant
growth processes in two dimensions. From statistical mechanics
point of view they describe interfaces in the continuum limits of
critical models. The relation of SLEs to the succesful conformal
field theory (CFT) approach of physicists has attracted attention since
the introduction of SLE. This paper gives a conformal field
theory interpretation of a generalization
SLE$(\kappa, \underline{\rho})$ of chordal SLEs, 
following the approach of e.g. \cite{Bauer:2002tf},
\cite{Bauer:2003vu}, \cite{Bauer:2004ij}, \cite{Bauer:2005jt}.
In a sense the aim is to show how the method of \cite{Bauer:2005jt}
can be applied to this case.
A different approach was presented in \cite{Cardy:2004xs}.

The chordal SLE$(\kappa)$ is known to be in some sense the most
general conformally invariant growth process in simply connected
regions that only depends on two points on the boundary of the
region, the starting point and the end point. SLE$(\kappa; \rho)$ is
the most general growth process depending on
three points, the starting point and two other points. For a higher
number of boundary points SLE$(\kappa; \underline{\rho})$ is no
longer the most general case.
Indeed, in section \ref{sec: CFT of SLE} we find a class of
such processes compatible with the CFT representation of statistical
physics models. SLE$(\kappa; \underline{\rho})$ appears as
the case where total charge needed in the Coulomb gas formalism
is zero and no screening charges are inserted.

Section \ref{sec: SLE basics} recalls the definition and
some basic properties of SLE$(\kappa; \underline{\rho})$.
For later application and in order to exhibit the role of
infinity, the SLE equation is written in a strip geometry
with another normalization of the uniformizing maps.
Section \ref{sec: CFT of SLE} contains the main steps of a
computation to show that a well chosen ratio of CFT
correlation functions becomes a martingale. The interpretation
and corresponding results for other SLEs are briefly discussed.
An example probability and it's possible meaning in the massless
free boson case $\kappa=4$ are the subject of
section \ref{sec: free field}. Finally, a more
algebraic approach (in the spirit of e.g. \cite{Bauer:2003kd})
to the result of this paper is sketched in section
\ref{sec: generating function}.

\section{SLE basics}
\label{sec: SLE basics}
\subsection{Definitions}
Let us give a brief introduction to SLEs and fix some notation.
For good general treatments of SLEs the reader may
want to see e.g. \cite{W:Lectures}, \cite{Kager:2004yd} or
\cite{Cardy:2005kh}.
The generalizations SLE$(\kappa; \rho)$ and
SLE$(\kappa; \underline{\rho})$ were introduced in
\cite{LSW:ConformalRestriction} and \cite{D:SLEduality}
respectively.

We will recall the definition of SLE$(\kappa; \underline{\rho})$
in the upper half plane $\bH = \{ z \in \bC \; : \; \im (z) > 0 \}$.
The parameters of the process are $\kappa \geq 0$, which corresponds
to how the curve looks locally (or the central charge of the CFT), and
$\underline{\rho} = (\rho_1, \ldots, \rho_n) \in \bR^n$ which has to
do with what kind of boundary conditions are imposed. The initial
conditions are the starting point of the curve $\xi_0 \in \bdry \bH = \bR$
and the locations of the boundary conditions $x_1, \ldots, x_n \in \bR$.

The chordal SLE$(\kappa, \underline{\rho})$ process in the upper
half plane is defined\footnote{Strictly
speaking one first assumes existence of a process $\xi_t$ with
the desired distribution and then considers the ODE
(\ref{eq: chordal SLE}).} by the L\"owners ordinary differential
equation for a uniformizing conformal map
\begin{equation}
\label{eq: chordal SLE}
\frac{\ud}{\ud t} g_t(z) = \frac{2}{g_t(z)-\xi_t} \textrm{ ,}
\qquad g_0 (z) = z \in \bH \textrm{ ,}
\end{equation}
where $\xi_t$, the driving process, is a solution of the
following It\^o stochastic differential equation
\begin{equation}
\label{eq: SLEkr driving process}
\ud \xi_t = \sqrt{\kappa} \; \ud B_t
    + \sum_{j=1}^n \frac{\rho_j}{\xi_t - g_t(x_j)} \ud t
\end{equation}
started at $\xi_0$ and $(B_t)_{t \geq 0}$ is a standard Brownian
motion.
The stochastic differential equation (\ref{eq: SLEkr driving process})
has a solution as long as all $|\xi_t - g_t(x_j)|$ stay bounded away
from zero, i.e. up to the
the time $\tau$ of first hitting of $\xi_t$ and some $g_t(x_j)$.
We can define $g_t$ by (\ref{eq: chordal SLE}) up to the same time.
For $z \in \bH$ the flow $g_t(z)$ is well defined up to the first
time $\tau_z$ that $g_t(z)$ hits $\xi_t$, i.e. $\tau_z =
\inf \{ t \in [0, \tau) : \liminf_{s \uparrow t} |g_s(z)-\xi_s| = 0 \}$.
The SLE hull at time $t < \tau$ is defined as
$K_t := \{ z \in \bH | \tau_z \leq t \}$.
Then $(K_t)_{0 \leq t < \tau}$ is a growing family of hulls,
$K_s \subset K_t$ for $s<t$.
The complement $\bH \setminus K_t$ is
simply connected and $g_t$ is the unique conformal mapping
$\bH \setminus K_t \rightarrow \bH$ with so called hydrodynamic
normalization
$g_t(z) = z + \order (1)$ at $z \rightarrow \infty$.
The key idea of SLE is to describe the random growth process of
hulls $(K_t)_{0 \leq t < \tau}$ in terms of the
family of uniformizing maps $(g_t)_{0 \leq t < \tau}$.

The hull grows only locally,
at the point $\gamma_t \in \overline{\bH}$, which is mapped
to the driving process, $g_t(\gamma_t) = \xi_t$.
The trace $t \mapsto \gamma_t$ is a continuos path almost surely.
For $\kappa \leq 4$ the trace is non-self-intersecting
and $K_t = \gamma_{[0,t]}$. For $4 < \kappa < 8$ the trace has
self-intersections and it ``swallows'' regions so
$\gamma_{[0, t]} \subsetneq K_t$.
In the parameter range $\kappa \geq 8$ the trace is space
filling, $\gamma_{[0, t]}$ has Hausdorff dimension $2$.

\bigskip

With $\rho = 0 \in \bR^n$ the process is called the chordal
SLE$(\kappa)$ from $\xi_0$ to $\infty$. The driving process
is then just a Brownian motion of variance parameter $\kappa$
and it can of course be defined for all $t \geq 0$,
that is $\tau = \infty$.

In \cite{Bauer:2005jt} it is shown how some multiple SLEs give
rise to SLE$(\kappa; \underline{\rho})$ processes.
SLE$(\kappa; \kappa - 6)$ is a chordal SLE$(\kappa)$
from $\xi_0$ to $x_1$, but the uniformizing maps $g_t$ have a
non standard normalization.
A special $n$SLE gives rise to SLE$(\kappa; \underline{\rho})$
with $\rho = (2, 2, \ldots , 2) \in \bR^{n-1}$.

\subsection{SLE$(\kappa; \rho)$ in a strip}
\label{sec: strip}
We can cast SLE$(\kappa; \underline{\rho})$ to a form which treats
$x_1$ and $\infty$ symmetrically. Assume that $x_1 < 0 = \xi_0$, other
cases lead to obvious changes. The conformal mapping
$m(z) = - \log \frac{-x_1}{z-x_1}$ from $\bH$ to
$\bS = \{ w \in \bC \; : \; 0 < \im (w) < \pi \}$ maps the
boundary points $x_1, 0 , \infty$ to $-\infty, 0, +\infty$.
Thus the family
\begin{eqnarray*}
h_t (w) = - \log \frac{\xi_t-g_t(x_1)}{g_t(m^{-1}(w))-g_t(x_1)}
\end{eqnarray*}
maps subsets $\bS \setminus m(K_t)$ to $\bS$, so it encodes
the growth of the conformal images of $K_t$. The normalization
of these conformal maps is such that $h_t(\pm \infty) = \pm \infty$
and $h_t(m(\gamma_t)) = 0$. A straightforward computation gives
\begin{eqnarray*}
\ud h_t(w) & = & -\sqrt{\kappa} \frac{\ud B_t}{\xi_t - g_t(x_1)} +
    \Big( - 3 + \frac{\kappa}{2} - \rho_1 - \sum_{j=2}^n \frac{\rho_j}{2} \\
& & + \coth (\frac{h_t(w)}{2}) + \sum_{j=2}^n \frac{\rho_j}{2}
    \coth (\frac{h_t(m(x_j))}{2}) \Big) \frac{\ud t}{(\xi_t - g_t(x_1))^2}
\textrm{ .}
\end{eqnarray*}
After a time change $\ud t = (\xi_t - g_t(x_1))^2 \ud s$ and
corresponding redefinition $\hat{h}_s = h_{t(s)}$ this takes a
more convenient form
\begin{eqnarray*}
\ud \hat{h}_s(w) & = & -\sqrt{\kappa} \; \ud B_s +
    \frac{\kappa - 6 - \sum_{j=1}^n \rho_j}{2} \; \ud s \\
& & + \sum_{j=1}^n \frac{\rho_j}{2} \coth (\frac{\hat{h}_s
    (\Tilde{x}_j)}{2}) \; \ud s + \coth (\frac{\hat{h}_s(w)}{2})
    \; \ud s
\textrm{ ,}
\end{eqnarray*}
where $\Tilde{x}_j = m(x_j)$, in particular $\Tilde{x}_1=-\infty$.
This suggests we should define
\begin{equation}
\rho_\infty = \kappa - 6 - \sum_{j=1}^n \rho_j
\end{equation}
and $\Tilde{x}_\infty = + \infty = m(\infty) \in \bdry \bS$.
The equation can also be written in terms of
\begin{eqnarray*}
g^\bS_s (w) & = & \hat{h}_s(w) + \eta_s \qquad \textrm{ and} \\
\ud \eta_s & = & \sqrt{\kappa} \; \ud B_s
    - \sum_{j \in \{ 1,2,\ldots,n, \infty \} } \frac{\rho_j}{2}
    \coth \big( \frac{g^\bS_s (\Tilde{x}_j) - \eta_s}{2} \big) \; \ud s
\end{eqnarray*}
so that $g^\bS_s$ uniformizes the complement of $m(K_{t(s)})$ in
$\bS$ and we have the L\"owner equation in the strip
\begin{eqnarray*}
\frac{\ud}{\ud s} g^\bS_s (w) = \coth \big( \frac{g^\bS_s(w)
    - \eta_s}{2} \big)
\textrm{ .}
\end{eqnarray*}

From the half plane equations for SLE$(\kappa; \underline{\rho})$ it
is not immediately clear what kind of boundary condition is imposed at
$\infty$. The above coordinate change to the strip $\bS$ shows the role
of $\infty$. In particular, for any SLE$(\kappa, \underline{\rho})$,
we have $\sum_{j=1}^n \rho_j + \rho_\infty = \kappa-6$, which in
section \ref{sec: CFT of SLE} is seen to be related to charge
neutrality in the Coulomb gas formalism.

As an example of the coordinate change we observe that the
SLE$(\kappa; \rho_1)$ which treats $x_1$ and $\infty$ symmetrically
is at the value $\rho_1 = (\kappa-6)/2$, since only in this case we
have $\rho_1 = \rho_\infty$. In the
strip geometry this special case is the dipolar SLE with its usual
normalization (see e.g. \cite{Bauer:2004ij}) --- the drift
$\frac{\rho_1}{2} \coth (-\infty) + \frac{\rho_\infty}{2}
\coth (+\infty)$ of the driving process $\eta_s$ vanishes.

\section{CFTs of SLE$(\kappa; \rho)$}
\label{sec: CFT of SLE}
In this section we find the conformal field theory appropriate for
SLE$(\kappa; \underline{\rho})$.
The result is a straightforward application of the ideas of
\cite{Bauer:2005jt} and \cite{Bauer:2004ij}, \cite{Bauer:2003vu},
\cite{Bauer:2002tf} to the present setup.
It turns out that we only need to use boundary primary fields as
boundary changing operators at the extra points. Not all such cases
are described by SLE$(\kappa; \underline{\rho})$, and this leads
us to propose a generalization.

\subsection{Statistical mechanics and martingales for SLEs}
We briefly remind the reader of a general argument about how SLEs
are related to statistical mechanics. The argument is presented in
more detail in \cite{Bauer:2005jt} and \cite{Bauer:2004ij}.

Different SLEs should represent interfaces in the continuum limits of
critical models of statistical mechanics with different boundary
conditions. We assume the model is defined in a simply connected
domain $D \subset \bC$ and we will use a parametrization of the
interface by a path $\gamma: [0, \infty) \rightarrow D$.
In two dimensions such
models have continuum limits desribed by conformal field theories.
The expected value of an observale $\Oper$ then becomes
\begin{eqnarray*}
\expect [ \Oper ] =
    \frac{\bra \Oper \; \Oper_{\mathrm{b.c.}}\ket^{\mathrm{CFT}}_D}
    {\bra \Oper_{\mathrm{b.c.}} \ket^{\mathrm{CFT}}_D}
\textrm{ ,}
\end{eqnarray*}
where $\bra \cdots \ket^{\mathrm{CFT}}_D$ denote conformal field
theory correlation functions in the domain $D$ of the model and
we have explicitly put operator $\Oper_{\mathrm{b.c.}}$ which
accounts for the boundary conditions of the model (one shouldn't
simply take vacuum expected value of $\Oper$). Note that it is
also necessary to divide by the correlation function of boundary
changes alone, which plays the role of the partition function ---
otherwise the expected value of the identity operator would differ
from unity.

Taking instead conditional expected values conditioned on the
knowledge of a portion of the interface $\gamma |_{[0,t]}$
defines a martingale (assuming integrability of the random
variable $\Oper$). But in many interesting cases the remaining
part of the interface arises from the same model defined in a
subdomain $D_t \subset D$, which is essentially the original
domain minus the portion of the interface.
In such a case 
\begin{equation} \label{eq: statmech martingale}
\expect \big[ \Oper \big| \gamma |_{[0,t]} \big] =
\frac{\bra \Oper \; \Oper_{\mathrm{b.c.}} (\gamma_t, x_1, x_2, \ldots)
    \ket^{\mathrm{CFT}}_{D_t}}
    {\bra \Oper_{\mathrm{b.c.}} (\gamma_t, x_1, x_2, \ldots)
    \ket^{\mathrm{CFT}}_{D_t}}
\end{equation}
is a martingale (the boundary condition changes only by
the location where the interface continues that is $\gamma_0$
is replaced with $\gamma_t$).
Using the transformation properties of CFT operators under
conformal mappings we can express this as a ratio of CFT
correlation functions in the original domain. SLEs give us
explicitly such transformations $g_t$ which map certain connected
component $D_t$ of $D \setminus \gamma_{[0,t]}$ back to the
original domain. By conformal covariance one usually chooses
$D = \bH$ when discussing SLEs.

Several cases involving different SLEs and different boundary
conditions have already been studied. The first
observation in \cite{Bauer:2002tf} was that the ordinary chordal
SLE corresponds to a theory of central charge
$c = c(\kappa) = \frac{(6-\kappa)(3\kappa-8)}{2\kappa}$ and a
boundary changing operator
$\Oper_{\mathrm{b.c.}} = \Psi_{1,2}(\infty) \Psi_{1,2}(0)$,
which creates the two ends of the interface $\gamma$ at $0$ and
$\infty$. The operator $\Psi_{1,2}$ stands for a boundary primary
field of a degenerate conformal weight\footnote{Here and in the
sequel we use the following Kac labeling of the conformal weights
$$
h_{r,s} = \frac{1}{16 \kappa} \big( \kappa^2 (r^2-1) - 8 \kappa (rs-1)
    + 16(s^2-1) \big) \textrm{ ,}
$$
which is convenient in the SLE context. These are the weights for
which the Verma module is reducible.}
$h_{1,2} = \frac{6-\kappa}{2\kappa}$ and this operator
has a vanishing descendant at level $2$. In all other cases the
central charge is given by the same formula and the interface is
again created by $\Psi_{1,2}$.
In \cite{Bauer:2003vu} the radial SLE was worked out
with the result $\Oper_{\mathrm{b.c.}} = \Phi_{0,\half} (z^*)
\Psi_{1,2} (0)$, $z^*$ being the endpoint of the
interface in the interior of the domain. The dipolar SLE, already
seen to be the symmetric
case of SLE$(\kappa; \rho)$, was studied in \cite{Bauer:2004ij}
and it has boundary conditions
$\Oper_{\mathrm{b.c.}} = \Psi_{0, \half} (x_-)
\Psi_{0,\half} (x_+) \Psi_{1,2}(0)$. The proposal for multiple SLEs
in \cite{Bauer:2005jt} was taken to include only creations of
interfaces on the real axis and something consistent at infinity:
$\Oper_{\mathrm{b.c.}} = \Psi_{h_\infty} (\infty)
\Psi_{1,2}(x_1) \cdots \Psi_{1,2} (x_n)$.
We will soon see that SLE$(\kappa; \underline{\rho})$ can be
understood with
\begin{eqnarray*}
\Oper_{\mathrm{b.c.}} = \Psi_{\delta_\infty} (\infty) \Psi_{\delta_1} (x_1)
    \cdots \Psi_{\delta_n} (x_n) \Psi_{1,2} (0)
\end{eqnarray*}
again including the creation of the interface at $0$ and boundary changes
at $x_1, \ldots, x_n, \infty$. The conformal weights are explicitly
\begin{eqnarray}
\label{eq: conformal weights}
\delta_j & = & \frac{\rho_j (\rho_j + 4 - \kappa)}{4 \kappa}
    \quad \textrm{ for $j=1, \ldots , n, \infty$.}
\end{eqnarray}
This formula fits nicely to the Coulomb gas formalism (the Coulomb
gas formalism of CFT is described in e.g \cite{DMS:CFT} and
\cite{Felder:1988zp}). To obtain
central charge $c = c(\kappa)$ one introduces a background charge
$-2 \alpha_0 = \frac{4-\kappa}{2 \sqrt{\kappa}}$. Then vertex
operators of charge $\alpha$ have conformal weight
$h(\alpha) = \alpha^2 - 2 \alpha_0 \alpha = \frac{1}{4 \kappa}
(2\sqrt{\kappa} \alpha) (2\sqrt{\kappa} \alpha + 4 - \kappa)$.
This means that the conformal weight corresponding to $\rho_j$ is
obtained with the charge $\alpha_j = \frac{\rho_j}{2 \sqrt{\kappa}}$.
There is, in addition, the operator creating the interface which
has charge $\alpha_{1,2} = \frac{1}{\sqrt{\kappa}}$. For
SLE$(\kappa, \underline{\rho})$ the sum of all charges vanishes
\begin{eqnarray*}
-2 \alpha_0 + \sum_{j \in \{1,2,\ldots,n,\infty\} } \alpha_j
    + \alpha_{1,2} = 0
\textrm{ ,}
\end{eqnarray*}
which means that there is no need for screening charges in the
Coulomb gas correlation function
\begin{eqnarray}
\nonumber
& & \Bra V_{\alpha_\infty} (\infty) V_{\alpha_1}(x_1) \cdots
    V_{\alpha_n}(x_n) V_{\alpha_{1,2}} (\xi) \Ket  \\
\nonumber
& = & \bra v^*_{2 \alpha_0 - \alpha_\infty} \; , \; V_{\alpha_1}(x_1)
    \cdots V_{\alpha_n}(x_n) V_{\alpha_{1,2}} (\xi) \; v_0 \ket \\
\label{eq: factorizable correlator}
& = & \Big( \prod_{j=1}^n
    (x_j-\xi)^{\rho_j/\kappa} \Big) \Big( \prod_{1 \leq j < k \leq n}
    (x_k - x_j)^{\rho_j \rho_k / 2 \kappa} \Big)
\textrm{ .}
\end{eqnarray}


\subsection{SLE with primary boundary changing operators}
We now consider the question of what kinds of SLEs can arise from creation
of one interface with $\Psi_{1,2}$ and with a number of boundary
changes implemented by boundary primary fields. All of these ``SLEs''
seem natural and probably deserve furter study. The
SLE$(\kappa; \underline{\rho})$ processes are the simplest such
processes (in the sense that one doesn't need screening charges)
and in this note we concentrate on them.

Take the domain to be the upper half-plane, $D = \bH$, and suppose
$\Oper_{\mathrm{b.c}} = \Psi_{\delta_\infty} (\infty)
\Psi_{\delta_n} (x_n) \cdots \Psi_{\delta_1} (x_1) \Psi_{1,2} (0)$.
Then the L\"owner mapping $g_t: H_t \rightarrow \bH$ with standard time
parametrization satisfies
$\frac{\ud}{\ud t} g_t(z) = \frac{2}{g_t(z)-\xi_t}$. Take the driving
process to be of the form
$\ud \xi_t = \sqrt{\kappa} \; \ud B_t + f \; \ud t$ so that the
interface looks locally like SLE$(\kappa)$ and we should
expect $c=c(\kappa)=\frac{(6-\kappa)(3\kappa-8)}{2\kappa}$ again.
For simplicity start with an observable which is a product of boundary
primary fields, $\Oper = \Psi_{h_1} (y_1) \cdots \Psi_{h_m} (y_m)$
with $y_1, \ldots, y_m \in \bR$. From now on all CFT correlation
functions will be in $\bH$ so we omit the subscript and superscript.

Applying the conformal transformation $g_t$ to
(\ref{eq: statmech martingale}) we find that the ratio
\begin{eqnarray*}
\frac{\prod_{i=1}^m g_t'(y_i)^{h_i} \; \Bra \prod_{i=1}^m \Psi_{h_i}
    (g_t(y_i)) \; \Psi_{\delta_\infty} (\infty) \;
    \prod_{j=1}^n \Psi_{\delta_j} (g_t(x_j)) \big)
    \Psi_{1,2} (\xi_t) \Ket}{\Bra \Psi_{\delta_\infty} (\infty) \;
    \prod_{j=1}^n \Psi_{\delta_j} (g_t(x_j))
    \Psi_{1,2} (\xi_t) \Ket}
\end{eqnarray*}
should be a local martingale. Let us denote the numerator by $N_t$
and denominator by $D_t$. 

The following computation is similar to one in \cite{Bauer:2005jt}
so only main steps are given here. Making use of the null field
$(-2 \sL_{-2} + \frac{\kappa}{2} \sL_{-1}^2) \Psi_{1,2}$
and Ward identity we compute the It\^o derivatives
\begin{eqnarray}
\nonumber
& \ud N_t = \sD N_t \qquad 
    \quad \ud D_t = \sD D_t & \\
\label{eq: D operator}
& \sD = \ud t \; \big( \sum_{j=1}^n \frac{2 \delta_j}{(\xi - X^{(j)})^2}
    + f \; \partial_\xi \big) + \ud B_t \; (\sqrt{\kappa} \; \partial_\xi) &
\textrm{ .}
\end{eqnarray}
The It\^o derivative of the ratio $N_t / D_t$ should have no drift
\begin{eqnarray*}
\ud \big( \frac{N_t}{D_t} \big) & = & \big( f - \kappa
    \frac{\partial_\xi D}{D} \big) \;
    \partial_\xi \big( \frac{N_t}{D_t} \big) \; \ud t + \sqrt{\kappa} \;
    \partial_\xi \big( \frac{N_t}{D_t} \big) \; \ud B_t
\textrm{ ,}
\end{eqnarray*}
which leads to the requirement $f = \kappa \; \partial_\xi \log D_t$.

The above requirement gives a class of interesting processes
generalizing SLE$(\kappa; \underline{\rho})$. One
could start with any conformal block for the correlation function
$D = \bra \Psi_{\delta_\infty} (\infty) \; \prod_{j=1}^n \Psi_{\delta_j}
(x_j) \; \Psi_{1,2} (\xi) \ket $ and define the driving process
$\xi_t$ by
\begin{eqnarray*}
\ud \xi_t = \sqrt{\kappa} \; \ud B_t + \kappa \big( \partial_\xi
    \log D (\xi_t; g_t(x_1), \ldots, g_t(x_n)) \big) \; \ud t
\textrm{ .}
\end{eqnarray*}

Here we are content to remark that with conformal weights $\delta_j$
given by (\ref{eq: conformal weights}), as a consequence of the charge
neutrality the Coulomb gas correlation function
(\ref{eq: factorizable correlator}) provides us the nice factorizable
$D(\xi; x_1, \ldots, x_n)$. This choice of $D$ 
gives rise to the drift term of SLE$(\kappa; \underline{\rho})$
\begin{eqnarray*}
f = \kappa \; \partial_\xi \log \Bra V_{\alpha_\infty} (\infty)
    V_{\alpha_1}(x_1) \cdots V_{\alpha_n}(x_n) V_{\alpha_{1,2}} (\xi) \Ket
    = \sum_{j=1}^n \frac{\rho_j}{\xi - x_j}
\textrm{ .}
\end{eqnarray*}
Note that $f \sim \rho_j / (\xi-x_j)$ fixes the asymptotic behavior
of $D$ as $\xi \rightarrow x_j$ and fusion rule then allows no other
conformal weight of a primary field at $x_j$ than
(\ref{eq: conformal weights}). 

\bigskip 

Concerning the generalization of SLE$(\kappa; \underline{\rho})$
suggested above and computations in SLE in general we also emphasise
that even if the conformal weight of some boundary changing operator
coincides with a weight $h_{r,s}$, the operator may or may not have a
vanishing descendant. Such phenomenon is not new in the SLE context,
see e.g. \cite{Bauer:2002tf} and \cite{Bauer:2004my}.

\section{Free field and SLE$(4; \rho)$}
\label{sec: free field}
At $\kappa=4$ the conformal field theory has central charge
$c=1$ and SLE should correspond to ``level sets''
of the massless free boson field\footnote{There is yet
unpublished work by Scott Sheffield and Oded Schramm which
shows that chordal SLE$(4)$ is a discontinuity curve between
two levels.}.
The next sections give a few indications of how
SLE$(\kappa; \underline{\rho})$ fits into the picture.

The application of SLE$(\kappa; \underline{\rho})$
to the free field was considered in \cite{Cardy:2004xs}.
The physics was discussed in more depth, but let us for
illustration purposes state a conclusion from there.
Let the free field $\varphi$ have piecewise constant Dirichlet
boundary conditions with jumps at $\xi_0, x_1, \ldots, x_n$.
The jump at $\xi_0$ has the critical value
$\lambda^*=\frac{1}{\sqrt{4g}}$, such that we can trace the
discontinuity line of this jump.
At $x_j$ one can take a jump of any size $\lambda_j$. The conformal
weight of the corresponding operator is $\delta_j = g \lambda_j^2$,
in particular for the critical jump size this is $\frac{1}{4}$.
The partition function for such a free field is
\begin{eqnarray*}
Z \sim \Big( \prod_{j=1}^n
    (x_j-\xi)^{\sqrt{g} \lambda_j} \Big) \Big( \prod_{1 \leq j < k \leq n}
    (x_k - x_j)^{2 g \lambda_k \lambda_j} \Big)
\textrm{ ,}
\end{eqnarray*}
of the form (\ref{eq: factorizable correlator}) that corresponds to
SLE$(\kappa; \underline{\rho})$. In \cite{Cardy:2004xs} there is
also a discussion that points to the interpretation given next.

\subsection{Probability to be on the left of the trace}
In this section we compute a certain example probability for
SLE$(\kappa, \rho)$ with $\kappa > 4$ and observe that its
limit at $\kappa=4$ has a natural interpretation in terms
of free field boundary conditions. Computation of this
probability for the dipolar SLE appeared in
\cite{Bauer:2004ij}.

Recall from section \ref{sec: strip} that the
SLE$(\kappa; \rho)$ equation in $\bS$ is conveniently
written in terms of the mapping $\hat{h}_s$ which maps the tip of
SLE trace to $0$
\begin{eqnarray*}
\ud \hat{h}_s(w) = -\sqrt{\kappa} \; \ud B_s +
    \frac{\kappa - 6 - 2\rho}{2} \; \ud s +
    \coth \big( \frac{\hat{h}_s(w)}{2} \big) \; \ud s
\textrm{ .}
\end{eqnarray*}
Suppose we have an analytic function $F : \bS \rightarrow \bC$
such that $F (\hat{h}_s(w))$ is a $\bC$-valued martingale. The
drift of $F(\hat{h}_s)$ should vanish, which by It\^o's formula
means
\begin{eqnarray*}
0 & = & \frac{\kappa - 6 - 2\rho}{2} F'(\hat{h}) + \coth \big(
    \frac{\hat{h}}{2} \big) F'(\hat{h}) + \frac{\kappa}{2} F''(\hat{h})
\textrm{ .}
\end{eqnarray*}
Conversely, the solutions to this equation give local
martingales. Constants solve the equation, but the other linearly
independent solution is an integral function of
\begin{eqnarray*}
\big( \sinh (u/2) \big)^{-4/\kappa}
    \exp \big( \frac{6-\kappa+2\rho}{\kappa} u \big)
\textrm{ .}
\end{eqnarray*}
The real and imaginary parts of that analytic function are
harmonic local martingales for SLE$(\kappa; \rho)$ in $\bS$.

We assume that $(\kappa - 8)/2 < \rho < (\kappa - 4)/2$ so that
the integral function is finite at $\pm \infty$ and that $\kappa>4$
so that it is finite at $0$.
Then the integral function can be taken to be
\begin{eqnarray*}
F(w) & = & \int_{-\infty}^w \big( \sinh (u/2) \big)^{-4/\kappa}
    \exp \big( \frac{6-\kappa+2\rho}{\kappa} u \big) \; \ud u 
\textrm{ .}
\end{eqnarray*}
Let us set
\begin{eqnarray*}
P^l_{\kappa; \rho}(w) = 1 - \frac{\im F(w)}{\im F(+\infty)}
\textrm{ ,}
\end{eqnarray*}
which takes the value $1$ at $- \infty$ and $0$ on the positive real axis.
If a point $w \in \bS$ is swallowed, we have
$\lim_{s \uparrow \sigma_w} \hat{h}_s(w)=0$ where $\sigma_w$ is the explosion
time of $w$ for the L\"owner equation in the strip. If $w$ is on the
left (resp. right) of the hull, we have
$\lim_{s \uparrow \sigma_w} \hat{h}_s(w) = -\infty$ (resp. $+\infty$). Thus
the function $P^l_{\kappa; \rho}(w)$ gives the
probability of $w$ being on the left of the hull,
$P^l_{\kappa; \rho}(\hat{h}_s(w)) =
\expect [ \unit_{w \textrm{ on left}} \; \big| \; \hat{\gamma} |_{[0,s]} ]$.

\subsection{Free field boundary conditions}

As $\kappa \downarrow 4$ the bounded harmonic function
$P^l_{\kappa; \rho} (w)$ has a limit $p^l_\rho (w)$, which is a harmonic
SLE$(4; \rho)$-martingale.
It's boundary conditions are $0$ on the positive real axis
and $1$ on the negative real axis, so it describes the probability
that $w$ will be on the left of the trace at the time the trace
hits the the upper boundary of the strip. On the upper boundary
$w = x + i \pi$ the directional derivative of $p^l_\rho$ to
direction $e^{- i \rho \pi / 2}$ vanishes.

If SLE$(4; \rho)$ describes a ``level set'' of the free
field, then the boundary condition on the upper boundary seems to
correspond to the free field $\varphi$ having a vanishing directional
derivative to the direction $e^{-i \rho \pi / 2}$. The computation
only made sense if $-2 < \rho < 0$ whereas the piecewise constant
Dirichlet boundary conditions of \cite{Cardy:2004xs} don't have such
restriction.

The boundary changing operator at $0$ has the dimension
$\frac{1}{4}$ and it changes boundary condition between two
different Dirichlet boundaries. The operators at $-\infty, +\infty$
have dimensions $\frac{\rho^2}{16}$, $\frac{\rho_\infty^2}{16}$
and they change the direction of the vanishing derivative by an
angle $\frac{\pi}{2} \rho$, $\frac{\pi}{2} \rho_\infty$
respectively. The total change of angle is
$\frac{\pi}{2} (\rho + \rho_\infty) = -\pi$ which is consistent
with returning to Dirichlet boundary condition. The particularly
interesting case $\rho=-1$ (the dipolar SLE) corresponds to
Neumann boundary condition as discussed in \cite{Bauer:2004ij}.

For a general SLE$(4; \underline{\rho})$ we always have a
boundary changing operator $\Psi_{1,2}$ of dimension $\frac{1}{4}$
at $\xi$ so the curve is created by a jump of critical size $\lambda^*$
in the Dirichlet boundary conditions. The other boundary changing
operators have dimensions $\frac{1}{16} \rho_j^2$ such that
$\sum_{j \in \{ 1,\ldots , n, \infty\} } \frac{\pi}{2} \rho_j = -\pi$.

\section{Virasoro module valued martingale}
\label{sec: generating function}
The approach of section \ref{sec: CFT of SLE} as well as of
\cite{Bauer:2005jt} is slightly different from that of
\cite{Bauer:2002tf}, \cite{Bauer:2003vu}, \cite{Bauer:2003kd},
\cite{Bauer:2004ij}.
We have shown that certain CFT correlation functions are
martingales for SLEs.

A more algebraic approach is to encode the state of the SLE
as a vector in a highest weight module for the Virasoro algebra
(the CFT Hilbert space should consist of highest weight
representations of the Virasoro algebra with a common central
charge $c(\kappa)$).
We think of starting from the vacuum $| \Omega \ket$
in an irreducible module of highest weight $0$ and then
applying intertwining operators $\Psi_{1,2} (\xi_t)$ and
$\Psi_{\delta_j} (g_t(x_j))$ to create the appropriate boundary
conditions in $\bH$. Then we apply an operator $G_t$ which
implements the conformal transformation $g_t^{-1}$ in space of
states.

In the next section we will give main steps of computation to
show that the process
\begin{eqnarray*}
| M_t \ket = \frac{1}{D(\xi_t; g_t(x_1), \ldots )} G_t
    \Psi_{\delta_n} (g_t(x_n)) \cdots \Psi_{\delta_1} (g_t(x_1))
    \Psi_{1,2} (\xi_t) | \Omega \ket
\end{eqnarray*}
in a highest weight module for the Virasoro algebra is a
(vector valued) martingale.
Correlation functions of section \ref{sec: CFT of SLE} can be
written as $\bra u | M_t \ket$
for some constant vectors $\bra u |$ in the dual of the module.
The example of section \ref{sec: free field}
corresponds to insertion of bulk primary field of dimension $0$
at $z=m^{-1}(w) \in \bH$, that is
$\bra u | = \bra \delta_\infty | \Phi_{0} (z)$.

\subsection{The computation of $\ud | M_t \ket$}

We left the computation of the It\^o derivative
of $| M_t \ket$ to this section. The drift will be shown to
be zero so that $| M_t \ket$ is local martingale with values
in a module for the Virasoro algebra. There are three kinds
of cancellations in the drift term. Some cancellations reflect the
fact that $G_t$ implements the transformation $g_t^{-1}$. For a
cancellation of another kind it is again crucial to have a null
descendant of the intertwining operator $\Psi_{1,2}$ which creates
the interface. Finally, it is important to take into account
the change of the partition function --- the third cancellations
arise from the correct choice of the denominator
$D(\xi ; x_1, \ldots, x_n)$, see section \ref{sec: CFT of SLE}.

Construction of the operator $G_t$, which implements the conformal
transformation $g_t^{-1}$ of SLE, was treated in \cite{Bauer:2003kd}.
From the L\"owner equation (\ref{eq: chordal SLE}) it follows that
\begin{eqnarray*}
\ud G_t = G_t e^{\xi_t L_{-1}} (-2 L_{-2}) e^{-\xi_t L_{-1}} \; \ud t
\textrm{ .}
\end{eqnarray*}

A boundary intertwining operator $\Psi_\delta$ of
conformal weight $\delta$ is a family of linear mappings between
Virasoro modules $M^{(0)}$ and $M^{(\infty)}$
parametrized by a boundary point $x \in \bdry \bH = \bR$
such that the following intertwining relation holds
\begin{eqnarray*}
[L_n, \Psi_\delta (x)] = \big( x^{1+n} \partial_x
    + (1+n) \delta x^n \big) \Psi_\delta (x)
\end{eqnarray*}
This means in particular that $L_{-1}$ generates
translations,
\begin{eqnarray*}
e^{t L_{-1}} \Psi_\delta (x) e^{-t L_{-1}} = \Psi_\delta (x+t)
\textrm{ .}
\end{eqnarray*}
The most useful form of the intertwining relation for the present
setup is
\begin{eqnarray*}
[e^{t L_{-1}} L_n e^{-t L_{-1}} , \Psi_\delta (x)] =
    \big( (x-t)^{1+n} \partial_x + (1+n) \delta (x-t)^n \big)
    \Psi_\delta (x)
\textrm{ .}
\end{eqnarray*}

We assume that $M^{(0)} = M_{1,1}$ is the irreducible
module of highest weight $h_{1,1}=0$, $M^{(1)} = M_{1,2}$ is
the irreducible module of highest weight
$h_{1,2} = \frac{6-\kappa}{2\kappa}$ and
$M^{(2)}, \ldots, M^{(n)}, M^{(n+1)}$ are some highest weight
modules for the Virasoro algebra. Let
$\Psi_{1,2} (x) : M^{(0)} \rightarrow M^{(1)}$ and
$\Psi_{\delta_j} : M^{(j)} \rightarrow M^{(j+1)}$ for
$j=1,\ldots,n$
be intertwining operators of conformal weights $h_{1,2}$ and
$\delta_j$.

In order to reduce the notation denote $g_t(x_j) = X^{(j)}_t$ and
$\frac{\partial}{\partial X^{(j)}_t} = \partial^{(j)}$ and
$f = \sum_{j=1}^n \frac{\rho_j}{\xi_t - X^{(j)}_t}$.
Let us first compute
\begin{eqnarray*}
& & \ud \big( D(\xi_t ; X^{(1)}_t, \ldots, X^{(n)}_t) | M_t \ket \big) \\
& = & G_t e^{\xi L_{-1}} (- \ud t \; 2 L_{-2}) e^{-\xi L_{-1}} \prod_j
    \Psi_{\delta_j} (X^{(j)}_t) \Psi_{1,2} (\xi) | \Omega \ket \\
& & + G_t \big( \ud t \; \sum_{j=1}^n \frac{2}{X^{(k)}_t-\xi}
    \partial^{(k)} \big)
    \prod_j \Psi_{\delta_j} (X^{(j)}_t) \Psi_{1,2} (\xi) | \Omega \ket \\
& & + G_t \prod_j \Psi_{\delta_j} (X^{(j)}_t) \big( \ud B_t \; \sqrt{\kappa}
    \partial_\xi + \ud t \; f \partial_\xi + \ud t \; \frac{\kappa}{2}
    \partial_\xi^2 \big) \Psi_{1,2} (\xi) | \Omega \ket
\textrm{ .}
\end{eqnarray*}
Commute $e^{\xi L_{-1}} L_{-2} e^{-\xi L_{-1}}$ in the first term to
the right and observe the cancellations with the second term.
Using the null vector
\begin{eqnarray*}
(-2 L_{-2} + \frac{\kappa}{2} L_{-1}^2) | h_{1,2} \ket = 0
\end{eqnarray*}
in the form
\begin{eqnarray*}
e^{\xi L_{-1}} ( -2 L_{-2} + \frac{\kappa}{2} \partial_\xi^2)
e^{-\xi L_{-1}} \Psi_{1,2} (\xi) | \Omega \ket = 0
\end{eqnarray*}
we have another cancellation in the first and third terms. The rest
can be written as
\begin{eqnarray*}
\ud \big( D(\xi_t ; X^{(1)}_t, \ldots, X^{(n)}_t) | M_t \ket \big)
    = \sD \big( D(\xi_t ; X^{(1)}_t, \ldots, X^{(n)}_t) | M_t \ket \big)
\textrm{ ,}
\end{eqnarray*}
where $\sD$ is as in (\ref{eq: D operator}). Using again the facts
$\ud D_t = \sD D_t$ and $f = \kappa \partial_\xi \log D_t$, as in
section \ref{sec: CFT of SLE}, we see that the drift of
$| M_t \ket$ vanishes.

\section{Conclusions}
We exhibited the conformal field theory appropriate for
SLE$(\kappa; \underline{\rho})$ in the statistical mechanics
interpretation of \cite{Bauer:2005jt} and \cite{Bauer:2004ij}.
The result turned out to be particularly simple in the Coulomb gas
formalism. In the course of study we proposed other
generalizations of SLE$(\kappa)$ to the case where boundary
conditions depend on more than three boundary points.

\bigskip

\emph{Acknowledgements:} The author is grateful to Michel Bauer
and Denis Bernard for interesting discussions on SLEs and conformal
field theory and to the anonymous referees for useful suggestions.

\end{document}